\documentclass[letterpaper]{llncs}

\usepackage{makeidx}  
\usepackage[nolist]{acronym}
\usepackage{amssymb}
\usepackage{icomma}
\usepackage{color}
\usepackage{array}
\usepackage{multirow}
\usepackage{amsmath}
\usepackage{layouts}
\usepackage{booktabs}
\usepackage{mathptmx}
\usepackage{graphicx}
\usepackage[caption=false]{subfig} 
\graphicspath{{figs/}}
\usepackage{hyperref}
\usepackage{breakurl}
\usepackage{algorithm}
\usepackage[noend]{algpseudocode}
\usepackage{cleveref}
\usepackage{siunitx} 
\usepackage{comment}
\usepackage{ifthen}

\newcommand{\bq}{\begin{equation}}
\newcommand{\eq}{\end{equation}}
\newcommand{\bytes}{\mbox{bytes}}
\newcommand{\byte}{\mbox{byte}}
\newcommand{\second}{\mbox{s}}

\newcommand{\flop}{\mbox{flop}}
\newcommand{\flops}{\mbox{flops}}

\newcommand{\cycle}{\mbox{cy}}
\newcommand{\iter}{\mbox{it}}

\newcommand{\BIT}{\mbox{\byte/\iter}}
\newcommand{\FR}{\mbox{\flops/\mbox{row}}}
\newcommand{\BR}{\mbox{\byte/\mbox{row}}}

\newcommand{\BC}{\mbox{\byte/\cycle}}
\newcommand{\GBS}{\mbox{G\byte/\second}}
\newcommand{\MBS}{\mbox{M\byte/\second}}

\newcommand{\GFS}{\mbox{G\flop/\second}}

\newcommand{\GHZ}{\mbox{GHz}}

\newcommand{\BF}{\mbox{\byte/\flop}}

\newcommand{\GB}{\mbox{GB}}

\newcommand{\MB}{\mbox{MB}}

\newcommand{\rlm}{roof{}line model}
\newcommand{\rl}{roof{}line}
\newcommand{\Rlm}{Roof{}line model}
\newcommand{\Rl}{Roof{}line}
\newcommand{\likwid}{\texttt{LIKWID}}
\newcommand{\likwidperfctr}{\texttt{likwid-perfctr}}
\newcommand{\likwidbench}{\texttt{likwid-bench}}
\newcommand{\lmbench}{\texttt{lmbench}}

\newcommand{\JHcomm}[1]{{\color{blue}!! {#1} !!\color{black}} }

\newcommand{\CAcomm}[1]{{\color{cyan}!! {#1} !!\color{black}} }

\usepackage{xcolor}

\renewcommand{\Comment}[2][0.5\linewidth]{%
	\leavevmode\hfill\makebox[#1][l]{\textcolor{gray}{$->$~\footnotesize{#2}}}}

\graphicspath{{pics/}}

\begin{document}

\begin{acronym}[DVFS]
    \acro{AGU}{address generation unit}
    \acro{AVX}{advanced vector extensions}
    \acro{CA}{cache agent}
    \acro{CL}{cache line}
    \acro{CoD}{cluster-on-die}
    \acro{DCT}{dynamic concurrency throttling}
    \acro{DIR}{directory}
    \acro{DP}{double precision}
    \acro{EDP}{energy-delay product}
    \acro{DP}{double precision}
    \acro{ECM}{execution-cache-memory}
    \acro{ES}{early snoop}
    \acro{FMA}{fused multiply-add}
    \acro{FP}{floating-point}
    \acro{HA}{home agent}
    \acro{HS}{home snoop}
    \acro{LFB}{line fill buffer}
    \acro{LLC}{last-level cache}
    \acro{MC}{memory controller}
    \acro{MSR}{model specific register}
    \acro{NT}{non-temporal}
    \acro{NUMA}{non-uniform memory access}
    \acro{OSB}{opportunistic snoop broadcast}
    \acro{RAPL}{running average power limit}
    \acro{SIMD}{single instruction, multiple data}
    \acro{SKU}{stock keeping unit}
    \acro{SP}{single precision}
    \acro{SP}{single precision}
    \acro{SSE}{streaming SIMD extensions}
    \acro{TDP}{thermal design power}
    \acro{UFS}{Uncore frequency scaling}
\end{acronym}

\newif\ifblind
\blindfalse

\frontmatter          
\pagestyle{headings}  

\mainmatter              
%
\title{Understanding HPC Benchmark Performance on Intel Broadwell and Cascade Lake Processors}
\titlerunning{HPC Benchmark Performance}  
\ifblind
\else
\author{Christie L. Alappat\inst{1} \and Johannes Hofmann\inst{2} \and Georg Hager\inst{1} \and Holger Fehske\inst{3}  \and\\ Alan R. Bishop\inst{4} \and Gerhard Wellein\inst{1,2}}
\authorrunning{Christie L. Alappat et al.} 
	\tocauthor{Christie L. Alappat et al.}
\institute{Erlangen Regional Computing Center (RRZE), 91058 Erlangen, Germany
\and
Department of Computer Science, University of Erlangen-Nuremberg,\\91058 Erlangen, Germany
\and
Institute of Physics, University of Greifswald, 17489 Greifswald, Germany
\and
Theory, Simulation and Computation Directorate, Los Alamos National Laboratory,\\Los Alamos, USA}
\fi
\maketitle              

\begin{abstract}
  Hardware platforms in high performance computing are constantly
  getting more complex to handle even when considering multicore CPUs
  alone.  Numerous features and configuration options in the hardware
  and the software environment that are relevant for performance are
  not even known to most application users or
  developers. Microbenchmarks, i.e., simple codes that fathom a
  particular aspect of the hardware, can help to shed light on such
  issues, but only if they are well understood and if the results can
  be reconciled with known facts or performance models. The insight
  gained from microbenchmarks may then be applied to real applications
  for performance analysis or optimization.  In this paper we
  investigate two modern Intel x86 server CPU architectures in depth:
  Broadwell EP and Cascade Lake SP. We highlight relevant hardware
  configuration settings that can have a decisive impact on code
  performance and show how to properly measure on-chip and off-chip
  data transfer bandwidths. The new victim L3 cache of Cascade Lake
  and its advanced replacement policy receive due attention.
  Finally we use DGEMM, sparse matrix-vector multiplication, and the
  HPCG benchmark to make a connection to relevant application
  scenarios.
\end{abstract}

\section{Introduction}

Over the past few years the field of high performance computing (HPC) has received attention from different
vendors, which led to a steep rise in the number of chip architectures. 
All of these chips have different performance-power-price points,
and thus different performance characteristics. 
This trend is believed to continue in the future with more
vendors such as Marvell, Huawei, and Arm entering HPC and related fields with new
designs. 
Benchmarking the architectures to understand their characteristics 
is pivotal for informed decision making and targeted code optimization. 
However, with hardware becoming more diverse, proper 
benchmarking is challenging and error-prone due to wide variety 
of available but often badly documented tuning knobs and settings.

In this paper we explore two modern Intel server processors,
Cascade Lake SP and Broadwell EP, using 
carefully developed micro-architectural benchmarks, then 
show how these simple microbenchmark codes become relevant
in application scenarios. During the process we demonstrate the
different aspects of proper benchmarking like the importance 
of appropriate tools, the danger of black-box benchmark code,
and the influence of different hardware and system settings.
We also show how simple performance models can help to draw correct
conclusions from the data.

Our microbenchmarking results highlight the changes from the Broadwell
to the Cascade Lake architecture and their impact on the performance of
HPC applications. Probably the biggest modification in this respect was
the introduction of a new L3 cache design.

This paper makes the following relevant contributions:
\begin{itemize}
\item We show how proper microarchitectural benchmarking can be used
  to reveal the cache performance characteristics of modern Intel
  processors. We compare the performance features of two recent
  Intel processor generations and resolve inconsistencies in published
  data.
\item We analyze the performance impact of the change in the L3
  cache design from Broadwell EP to Skylake/Cascade Lake SP and investigate
  potential implications for  HPC applications
  (effective L3 size, prefetcher, scalability). 
\item We use standard kernels and benchmarks from selected application
  areas to investigate the characteristic performance change for these
  fields:
\item For DGEMM we show the impact of varying core and Uncore clock speed, problem size,
  and sub-NUMA clustering on Cascade Lake SP.
\item For a series of sparse matrix-vector multiplications we show
  the consequence of the nonscalable L3 cache and the benefit
  of the enhanced effective L3 size on Cascade Lake SP.
\item To understand the performance characteristics of the HPCG benchmark,
  we construct and validate the \rl\ model for all its components and the full solver
  for the first time. Using the model we identify an MPI desynchronization
  mechanism in the implementation that causes erratic performance
  of one solver component.
\end{itemize}
This paper is organized as follows. After describing the benchmark systems setup
in \Cref{sec:testbed}, microarchitectural analysis using microbenchmarks (e.g.,
load and copy kernels and STREAM) is performed in \Crefrange{sec:singlecore}{sec:multicore}\@.
In \Cref{sec:implications} we then revisit the findings and see how they affect
code from realistic applications. \Cref{sec:summary} concludes the paper.


\subsubsection{Related work}

Molka et al.~\cite{BenchIT} used their BenchIT microbenchmarking framework to
thoroughly analyze latency and bandwidth across the full memory hierarchy of
Intel Sandy Bridge and AMD Bulldozer processors, but no application
analysis or modeling was done.
Hofmann et al.~\cite{Hofmann2016,Hofmann:2017} presented microbenchmark results for
several Intel server CPUs. We extend their methodology towards Cascade Lake SP and
also focus on application-near scenarios.
Saini et al.~\cite{lmbNASA2016,lmbNASA2017} compared a range of Intel server processors
using diverse microbenchmarks, proxy apps and application codes. They did  not, however,
provide a thorough interpretation of the data in terms of the hardware architectures. 
McIntosh-Smith et al.~\cite{ARM_Bristol} compared the Marvell ThunderX2 CPU with
Intel Broadwell and Skylake using STREAM, proxy apps and full applications, but
without mapping architectural features to microbenchmark experiments.
Recently, Hammond et al.~\cite{lmbSANDIA2018,lmbSANDIA2019} performed a benchmark analysis of the
Intel Skylake and Marvell ThunderX2 CPUs, presenting data that was partly in contradiction
to known microarchitectural features. A deeper investigation of these issues formed
the seed for the present paper. 



\section{Testbed and environment} 
\label{sec:testbed}


All experiments were carried out on one socket each of Intel's Broadwell-EP (BDW)
and Cascade Lake-SP (CLX) CPUs. These represent the  previous and current
generation CPUs in the Intel line of architectures which represents more than
85\% of the \textsc{top500} list published in November 2019.
\Cref{tab:testbed} summarizes the key specifications of the testbed.

The Broadwell-EP architecture has a traditional Intel design with
a three-level inclusive cache hierarchy. The L1 and L2 caches are private to each core
and the L3 is shared. BDW supports the AVX2 instruction set, which is capable of
256-bit wide SIMD.  The Cascade Lake-SP
architecture has a shared non-inclusive victim L3 cache. The particular model
in our testbed supports the AVX-512 instruction set and has 512-bit wide SIMD.
Both chips support ``Cluster on Die [CoD]'' (BDW) or ``Sub-NUMA Clustering [SNC]'' (CLX) feature, by which
the chip can be logically split in two ccNUMA domains.

Both systems ran Ubuntu version 18.04.3 (Kernel 4.15.0).
The Intel compiler version 19.0 update 2 with 
the highest optimization flag (-O3) was used throughout.
Unless otherwise stated, we added  architecture-specific flags
\texttt{-xAVX} (\texttt{-xCORE-AVX512} \texttt{-qopt-zmm-usage=high}) for BDW
(CLX).
The \likwid\ tool suite in version 4.3 was used
for 
performance counter measurements (\likwidperfctr\ tool)
and benchmarking (\likwidbench\ tool). 
Note that \likwidbench\  generates assembly kernels 
automatically, providing full control over the finally executed code.

\begin{table*}[!tb]
	\centering\footnotesize
	\caption{\label{tab:testbed}Key specification of test bed machines.}
	\begin{tabular}{l c c c}
		\toprule
		Microarchitecture               & Broadwell-EP (BDW)           & Cascade Lake-SP (CLX)             \\
		\midrule
		Chip Model                      & Xeon E5-2697 v4               & Xeon Gold 6248                    \\
		Supported core freqs            & 1.2--3.6\,GHz                 & 1.2--3.9\,GHz                     \\
		Supported Uncore freqs          & 1.2--2.8\,GHz                 & 1.0--2.4\,GHz                     \\
		Cores/Threads                   & 18/36                         & 20/40                             \\
		Latest SIMD extension           & AVX2 / FMA                    & AVX-512                           \\
		L1 cache capacity               & 18$\times$32\,kB              & 20$\times$32\,kB                  \\
		L2 cache capacity               & 18$\times$256\,kB             & 20$\times$1\,MB                   \\
		L3 cache capacity               & 45\,MB (18$\times$2.5\,MB)    & 27.5\,MB (20$\times$1.375\,MB)    \\
		Memory Configuration            & 4 ch. DDR4-2400               & 6 ch. DDR4-2933                   \\
		LD/ST throughput           & 2 LD, 1 ST (AVX)                & 2 LD, 1 ST (AVX512)                         \\
		L1 - L2 bandwidth          & 64 B/cy                       & 64 B/cy                      \\
		L2 - L3 bandwidth          & 32 B/cy                  & 16 B/cy + 16 B/cy                     \\
		Theor. Mem. Bandwidth           & 76.8\,GB/s                    & 140.8\,GB/s                       \\
		Operating system                     & Ubuntu 18.04.3              & Ubuntu 18.04.3    \\ 
		Compiler	& Intel 19.0 update 2 & Intel 19.0 update 2 \\
		\bottomrule
	\end{tabular}
\end{table*}

\subsubsection{Influence of machine and environment settings}
\begin{figure}[!tb]
	\centering
	\hspace*{\fill}
	\includegraphics*[scale=0.70]{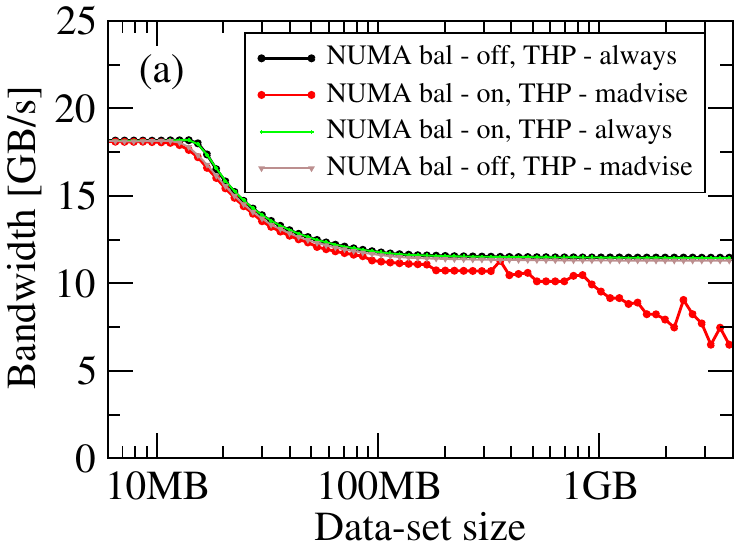}
	\hfill
	\includegraphics*[scale=0.70]{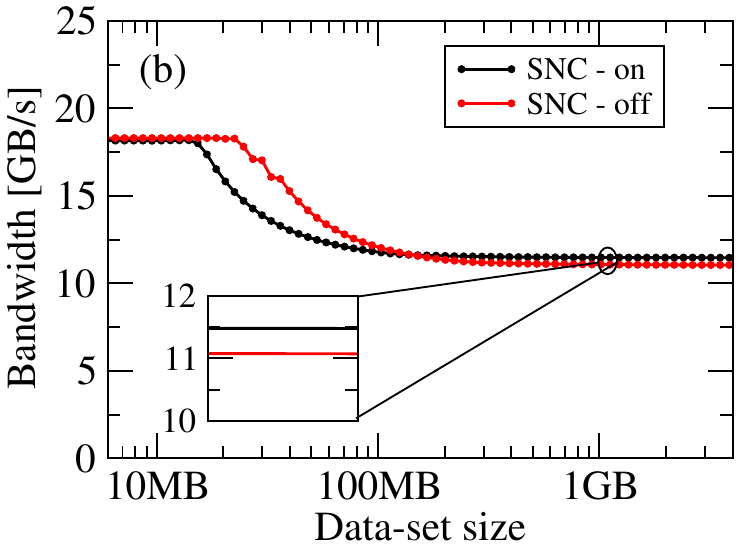}
	\hspace*{\fill}
	\caption{\label{fig:settings_impact} 
	  Performance impact of (a) NUMA balancing and transparent huge pages and (b) sub-NUMA clustering on a load-only
          streaming benchmark on CLX.
	}
\end{figure}

The machine and environment settings are a commonly neglected aspect
of benchmarking, although they can have a decisive impact on performance.
One should therefore take care to document all available settings.
\Cref{fig:settings_impact}(a) shows the influence of different operating 
system (OS) settings on a serial load-only benchmark running at 1.6\,\GHZ\ on CLX
for different data-set sizes in L3 and main memory.
With the default OS setting (NUMA balancing on 
and transparent huge pages (THP) set to ``madvise''), we can see a
2$\times$ hit in performance for big data sets.
This behavior also strongly depends on the OS 
version. We observed it with Ubuntu 18.04.3 (see \Cref{tab:testbed}).
Consequently, we use the setting that gives highest
performance, i.e., NUMA balancing off and THP set to ``always,'' for
all subsequent experiments.

Modern systems have an increasing number of knobs to tune on system startup.
\Cref{fig:settings_impact}(b) shows the consequences of the
sub-NUMA clustering  (SNC) feature on CLX for the load-only benchmark.
With SNC active the single core has local access to
only one sub-NUMA domain causing the shared L3 size to be halved. 
For accesses from main memory, disabling SNC slightly reduces
the single core performance by 4\% as seen in the inset of \Cref{fig:settings_impact}(b).

\section{Single-core bandwidth analysis}
\label{sec:singlecore}
\begin{figure}[!tb]
	\begin{minipage}{0.58\textwidth}
		\includegraphics[width=1\textwidth]{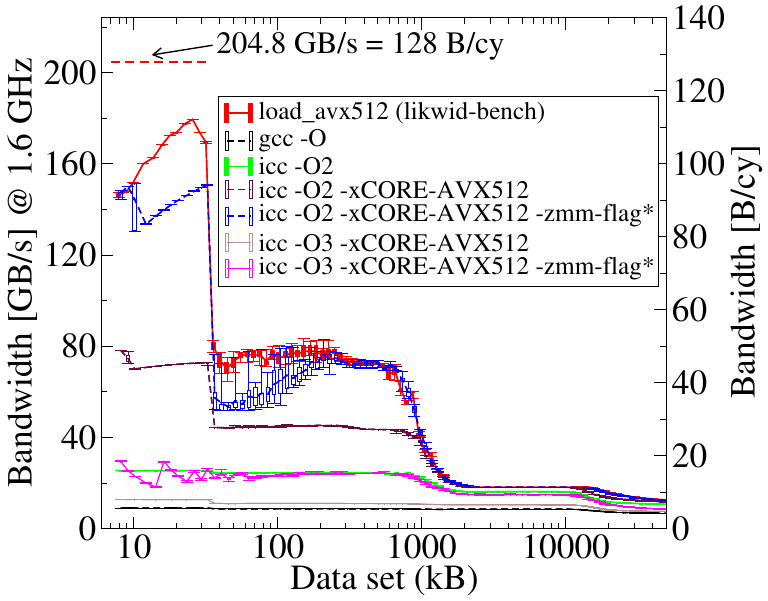}
	\end{minipage}
	\hspace{0.6em}
	\begin{minipage}{0.4\textwidth}
		\caption{\label{fig:likwid_vs_lmbench} Load only bandwidth as a function of 
			data set size on CLX. 
			The plot compares the bandwidth obtained from 
			\likwidbench\ with that of \lmbench. \likwidbench\ is able 
			to achieve 88\% of the theoretical L1 bandwidth limit (128 \BC). 
			The extreme sensitivity of \lmbench\ benchmark results to compilers and
			compiler flags is also shown. The ``zmm-flag*'' refers to 
			the compiler flag \texttt{-qopt-zmm-usage=high}.
		}
	\end{minipage}
\end{figure}

Single-core bandwidth analysis is critical to understand the 
machine characteristics and capability for a wide range of 
applications, but it requires great care especially when
measuring cache bandwidths since any extra cycle will directly 
change the result. To show this we choose the popular
bandwidth 
measurement tool \lmbench\ \cite{lmbench_tool}. 
\Cref{fig:likwid_vs_lmbench} shows the load-only (full-read or frd) 
bandwidth obtained by \lmbench\ as a function of data set size on CLX
at 1.6\,\GHZ.
Ten runs per size are presented in a box-and-whisker plot.

Theoretically, one core is capable of two AVX-512 loads per cycle 
for an L1 bandwidth of 128\,\BC\ (204.8\,\GBS\,@\,1.6\,\GHZ).
However, with the compiler option \verb.-O2. (default setting) it
deviates by a huge  factor of eight (25.5\,\GBS) 
from the theoretical limit. The characteristic strong 
performance gap between L1 and L2 is also missing.
Therefore, we tested different compiler
flags and compilers to see the effect (see \Cref{fig:likwid_vs_lmbench})
and observed a large span of performance values.
Oddly, increasing the level of optimization (\verb.-O2. vs \verb.-O3.) 
dramatically decreases the performance. The highest bandwidth was attained
for \verb.-O2. with the architecture-specific flags mentioned in \Cref{sec:testbed},
but even then the run-to-run variability is high, especially in
L2.

It is impossible to draw any profound conclusions 
about the machine characteristic from such measurements, thus
\lmbench\ results for frd (e.g.,
\cite{lmbNASA2016,lmbNASA2017,lmbSANDIA2018,lmbSANDIA2019}) should
be interpreted with due care.
However, employing  proper tools one can attain bandwidths close
to the limits. 
This is demonstrated by the AVX-512 load-only
bandwidth results obtained using \likwidbench\ \cite{Treibig:2011:3}.
As seen in \Cref{fig:likwid_vs_lmbench} with \likwidbench\
we get 88\% of the theoretical limit in L1, the expected
drops at the respective cache sizes, and much less run-to-run
fluctuation.

\begin{figure}[tb]\centering
	\includegraphics*[width=0.85\textwidth]{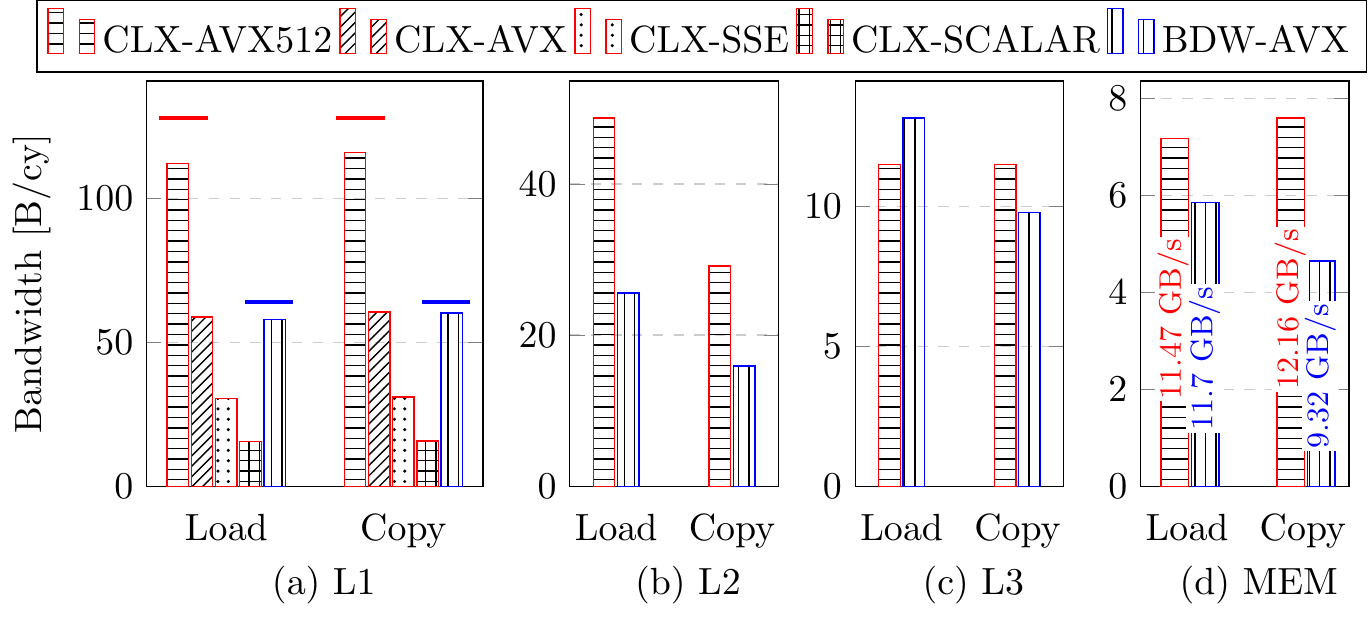}
	\caption{\label{fig:cache_bw_likwid} 
		Single-core bandwidth measurements in all memory hierarchy levels
		for load-only and copy benchmarks (\likwidbench). The bandwidth is
		shown in \BC, which is a frequency-agnostic unit for L1 and L2 cache.
		For main memory the bandwidth in \GBS\ at the base AVX512/AVX 
		clock frequency of 1.6\,\GHZ/2\,\GHZ\ for CLX/BDW is also indicated.
		Different SIMD widths are shown for CLX in L1.
		Horizontal lines denote theoretical bandwidth limits.
	}
\end{figure}

\Cref{fig:cache_bw_likwid} shows application bandwidths\footnote{Application bandwidth refers to the bandwidth as seen by the
application without the inclusion of hidden data traffic like write-allocate transfers.}
from different memory hierarchy levels of BDW and CLX (load-only and copy kernels).
The core clock frequency was fixed at 1.6 and 2\,\GHZ\ for CLX and BDW, respectively, with SNC/CoD switched on.
The bandwidth is shown in \BC, which makes it independent of core clock
speed for L1 and L2 caches.
Conversion to \GBS\ is done by multiplying the \BC\ value with
the clock frequency in \GHZ. The effect of single-core  L1 bandwidth
for scalar and different SIMD width is also shown in \Cref{fig:cache_bw_likwid}(a) for CLX. 
It can be seen that the bandwidth reduces by 2$\times$
as expected when the SIMD width is halved each time.

\section{Intel's new shared L3 victim cache}
\label{sec:L3}
From BDW to CLX there are no major observable changes to the behavior of
L1 and L2 caches, expect that the L2 cache size has been significantly
extended in CLX. However, starting from Skylake (SKX) the L3 cache
has been redesigned.
In the following  we study the effect of this newly designed
noninclusive victim L3 cache.

\subsection{L3 cache replacement policy}
\label{sec:l3_replacement}

A significant change with respect to the L3 cache concerns its
replacement policy. Prior to SKX, Intel processors' L3 caches implemented a
so-called pseudo-LRU cache-replacement policy~\cite{ia32opt:2016}. This policy
was fixed in the sense that the same pseudo-LRU policy was applied to all
workloads. Starting with SKX, this situation changed. Empirical data indicates
that SKX's L3 cache uses an adaptive replacement 
policy~\cite{Qureshi:2007}. This means the
processor implements a set of different replacement policies. During execution
the constantly processor selects a suitable policy depending on the current
workload in an attempt to maximize the L3-cache hit rate.

Experimental analysis suggests that the replacement policy 
selected by the
processor for streaming-access patterns involves placing new cache lines only
in one of the eleven ways of each cache set---the same strategy that is used
when prefetching data using the \texttt{prefetchnta} instruction (cf. Sect.
7.6.2.1 in~\cite{ia32opt:2016}). Consequently, data in the remaining ten ways
of the sets will not be preempted and can later be reused.

\begin{figure}[tb]
    \centering
    \hspace*{\fill}
	\includegraphics*[scale=0.70]{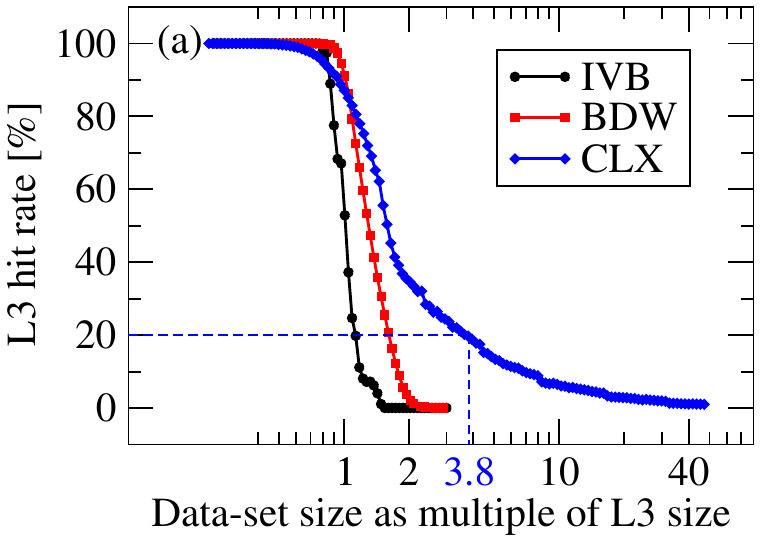}
    \hfill
    \includegraphics*[scale=0.70]{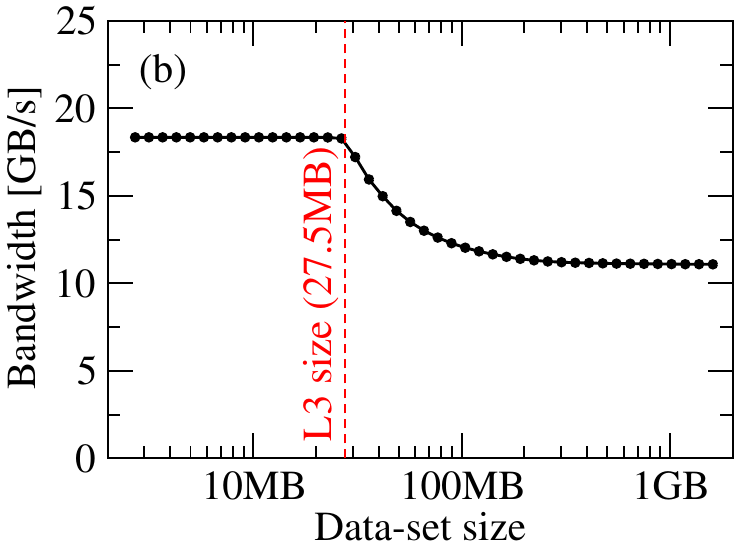}
    \hspace*{\fill}
    \caption{Demonstration of the implications of the new cache-replacement
    policy of CLX using L3-cache (a) hit rate and (b) bandwidth for a load-only
    data-access pattern. 
    In (a), the data for the older
    	 Intel Ivy Bridge Xeon E5-2690\,v2 (IVB) processor is shown as well.}
    \label{fig:L3-repl-policy}
\end{figure}

Figure~\ref{fig:L3-repl-policy}a demonstrates the benefit of this replacement
policy optimized for streaming-access patterns by comparing it to the
pseudo-LRU policies of previous generations' L3 caches. The figure shows the
L3-cache hit rate\footnote{Based on performance-counter data for the
\texttt{MEM\textunderscore{}\-LOAD\textunderscore{}\-RETIRED\textunderscore{}\-L3\textunderscore{}\-HIT}
and \texttt{MISS} events.} for different data-set sizes on different
processors for a load-only data-access pattern. To put the focus on the impact
of the replacement policies on the cache hit rate, hardware prefetchers were
disabled during measurements.  Moreover, data-set sizes are normalized to
compensate the processors' different L3-cache capacities. The data indicates
that older generations L3 caches offer no data reuse for data sets two times
the cache capacity, whereas CLX's L3 delivers hit rates of 20\% even for data
sets almost four times its capacity. Even for data more than ten times the L3
cache's size can reuse be detected on CLX.

The fact that this improvement can also be observed in practice is
demonstrated in Figure~\ref{fig:L3-repl-policy}b, which shows measured
bandwidth for the same load-only data-access pattern on CLX. For this
measurement, all hardware prefetchers were enabled. The data indicates that
the L3-cache hit-rate improvements directly translate into higher-than-memory
bandwidths for data sets well exceeding the L3 cache's capacity.

\subsection{L3 scalability}
\label{sec:l3_scalability}
Starting from Intel's Sandy Bridge architecture (created in 2011) 
 the shared L3 cache of all the Intel architectures up to Broadwell 
 is known to scale very well with the number of cores \cite{Hofmann:2017}. 
However, with SKX onwards the L3 cache architecture has changed from 
the usual ring bus architecture to a mesh architecture.
 Therefore in this section we test the scalability of this new L3 cache.
 
In order to test the L3 scalability 
 we use again the \likwidbench\ tool and run the benchmark with
increasing number of cores.
The data-set size was carefully chosen to be 2\,\MB\ per core 
to ensure that the size is sufficiently bigger than the L2 cache 
 however small enough such that no significant data traffic is 
 incurred from the main memory. 
 
 \begin{figure}[!tb]
 	\centering
 	\subfloat[Load]{\label{fig:L3_scale_load}
 		\includegraphics*{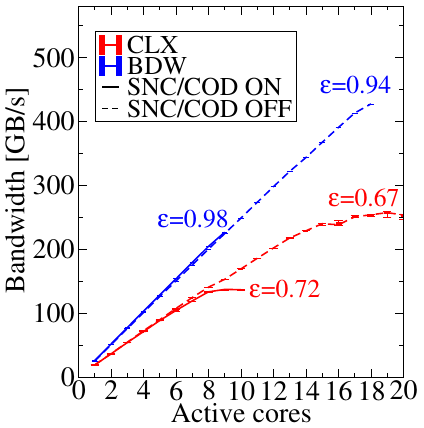}}
 	\subfloat[Copy]{\label{fig:L3_scale_copy}
 		\includegraphics*{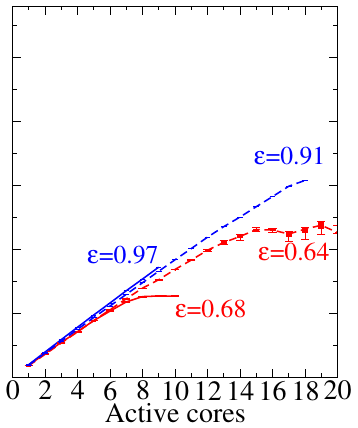}}
 	\subfloat[Update]{\label{fig:L3_scale_update}
 		\includegraphics*{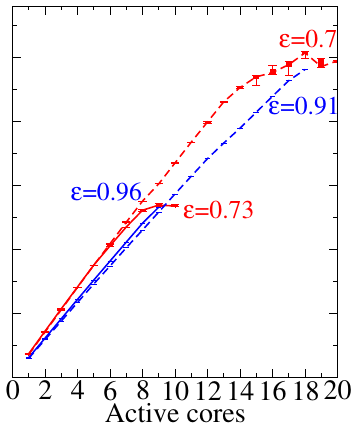}}
 	
 	\caption{\label{fig:L3_scaling} L3 Bandwidth of load, copy and update benchmark
 		measured on CLX and BDW. The saturation of L3 bandwidth on
 		CLX architecture can be clearly seen. The parallel efficiency
 		of each NUMA domain is further labeled in the plot.}
 \end{figure}

 
 The application bandwidths of the three basic kernels 
  load-only, copy and update are shown in \Cref{fig:L3_scaling} for 
  CLX\footnote{The result on SKX is exactly the same as CLX} and BDW.
 As the update kernel has equal number of loads and stores it
 shows the maximum attainable performance on both architectures. 
 Note that also within cache hierarchies write-allocate transfers
occur leading to lower copy application bandwidth. 
The striking difference between CLX and BDW for load-only bandwidth
can finally be explained by the bi-directional L2-L3 link 
on CLX which  only has half the load-only bandwidth as BDW (see \Cref{tab:testbed}).

 In terms of scalability we find that the BDW scales almost linearly 
 and attains an efficiency within 90\%, proving that the BDW has almost
 purely scalable L3 cache.
However, with CLX this behavior has changed drastically and the L3 cache 
saturates at higher core counts
both with and without SNC enabled, yielding an efficiency of about 70\%.
Due to this for the applications that employ L3 cache blocking it might 
be worthwhile to investigate the impact of switching to pure L2 blocking
 on SKX and CLX architectures.
In case of applications that use the shared property of L3 cache
like some of the temporal blocking schemes \cite{Wellein_wavefront}
a similar saturation effect as seen in \Cref{fig:L3_scaling} might be visible.

The effect of SNC/COD mode is also shown in \Cref{fig:L3_scaling}, with dotted lines corresponding
to SNC off mode and solid to SNC on mode. For CLX with SNC off mode the bandwidth 
attained at half of the socket (ten threads) is higher than SNC on mode.
This is due to the availability of $2 \times$ more L3 tiles and 
controllers with SNC off mode.


\section{Multi-core scaling with STREAM}
\label{sec:multicore}

\begin{figure}[tb]
    \centering
    \subfloat[CoD/SNC enabled]{\label{fig:triad_CoD}
    \includegraphics*[width=0.4\linewidth]{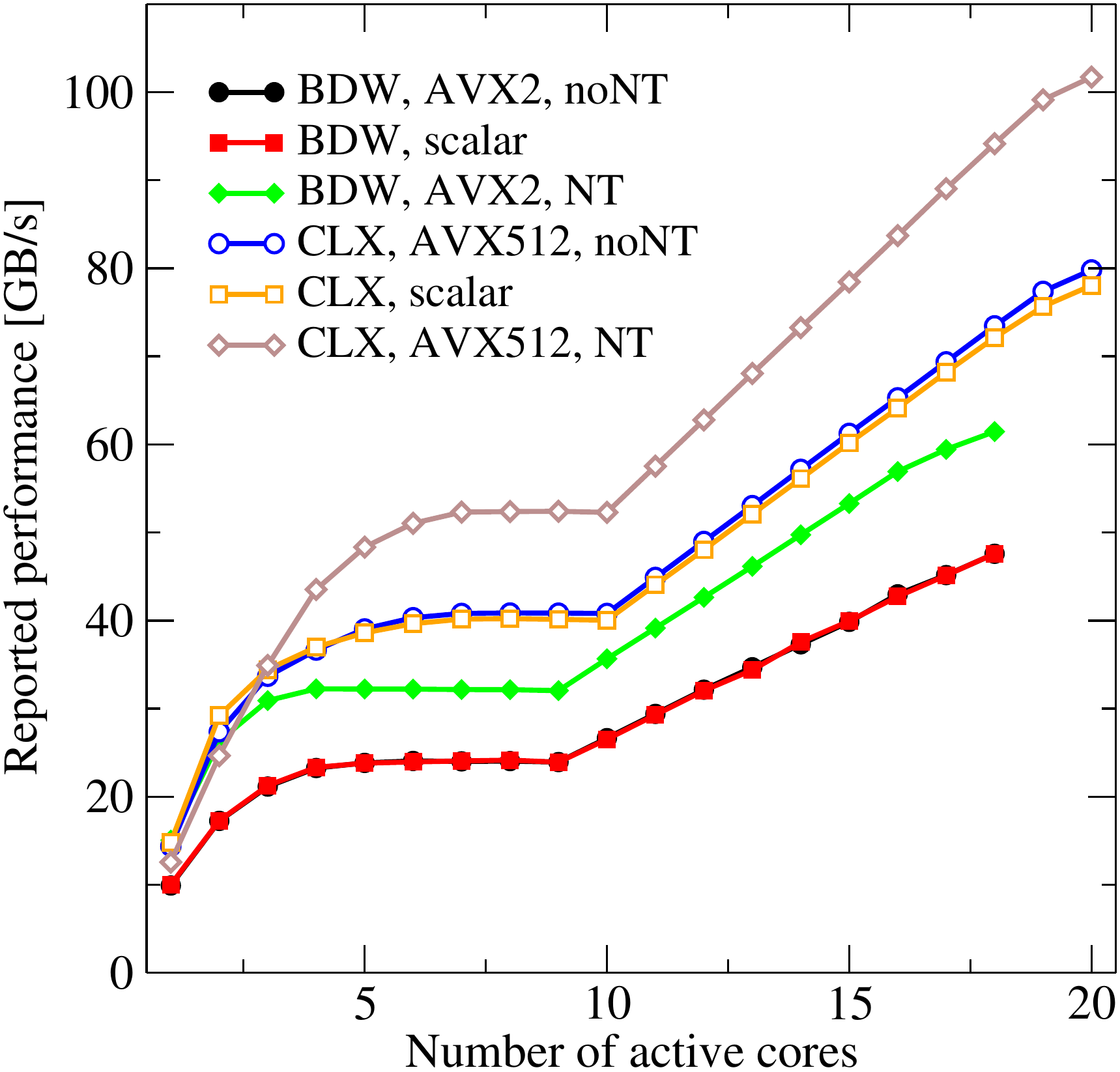}}\qquad
    \subfloat[CoD/SNC disabled]{\label{fig:triad_noCoD}
    \includegraphics*[width=0.4\linewidth]{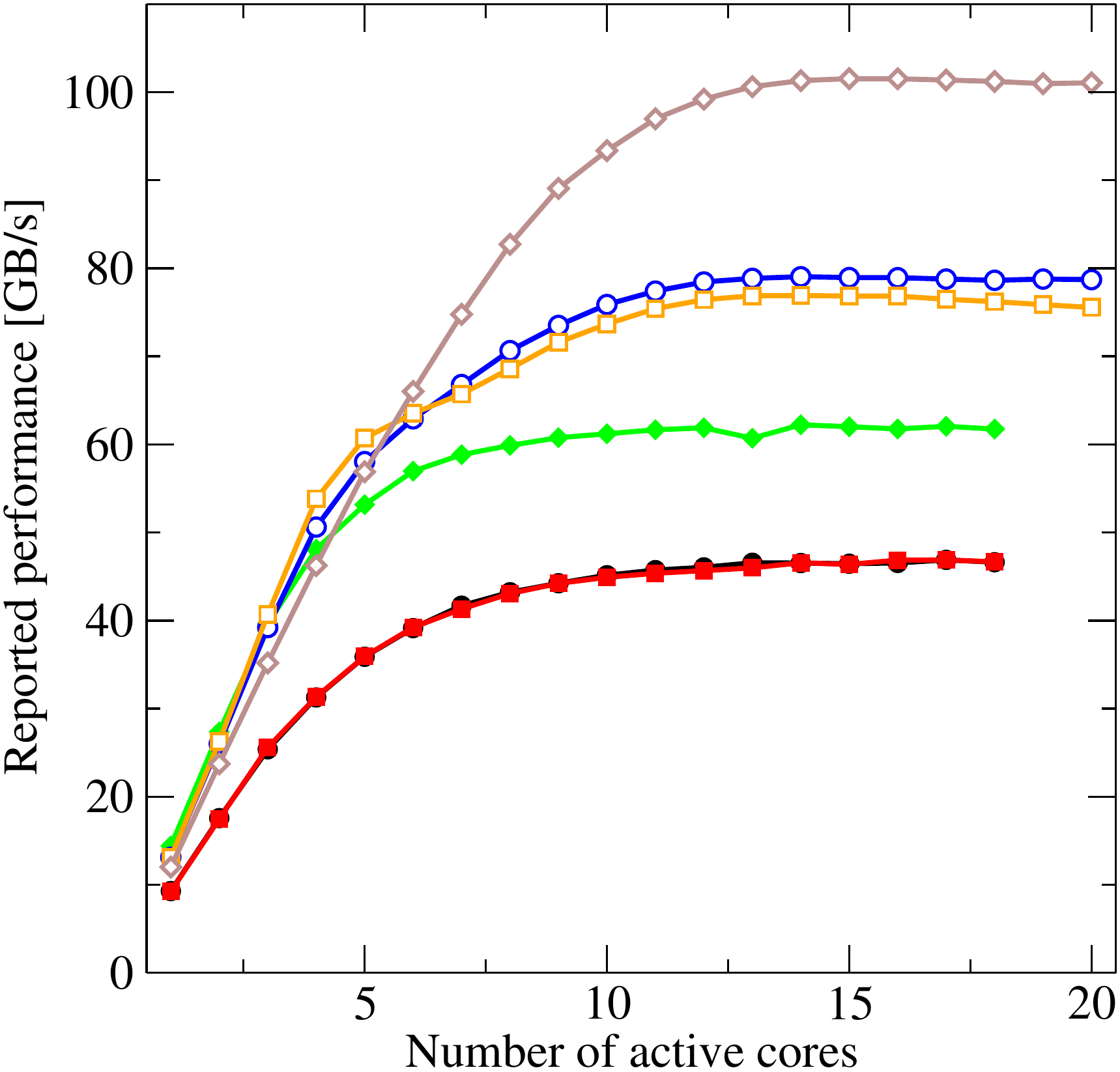}}
    \caption{\textsc{STREAM triad} scaling on BDW (closed symbols) and
      CLX (open symbols) with (a) CoD/SNC enabled and (b) CoD/SNC
      disabled. ``NT'' denotes the use of nontemporal stores (enforced
      by the \texttt{-qopt-streaming-stores always}), with ``noNT''
      the compiler was instructed to avoid them (via
      \texttt{-qopt-streaming-stores never}), and the ``scalar''
      variant used non-SIMD code (via \texttt{-no-vec}).  The working
      set was 2\,\GB. Affinity was enforced across physical cores,
      filling the socket from left to right.}
    \label{fig:STREAM}
\end{figure}
The STREAM benchmark~\cite{McCalpin:1995} has the purpose of measuring
the achievable memory bandwidth of a processor. Although the code
comprises four different loops, their performance is generally
similar and usually only the triad (\verb.A(:)=B(:)+s*C(:).) is reported.
The benchmark output is a bandwidth number in \MBS, assuming 24\,\byte\
of data traffic per iteration. The rules state that the working set size should
be at least four times the LLC size of the CPU\@. In the light of the new
LLC replacement policies (see \Cref{sec:l3_replacement}), this appears too small
and we chose a 2\,\GB\ working set for our experiments.

Since the target array \verb.A. causes write misses, the assumption
of the benchmark about the code balance is wrong if write-back caches
are used and write-allocate transfers cannot be avoided. X86 processors
feature \emph{nontemporal store} instructions (also known as
\emph{streaming stores}), which bypass the normal cache hierarchy
and store into separate write-combine buffers. If a full cache line
is to be written, the write-allocate transfer can thus be avoided.
Nontemporal stores are only available in SIMD variants on Intel processors,
so if the compiler chooses not to use them (or is forced to by a directive
or a command line option),  write-allocates will occur and the memory bandwidth
available to the application is reduced. This is why vectorization
\emph{appears} to be linked with better STREAM bandwidth, while it is actually
the nontemporal store that cannot be applied for scalar code. 
Note that a careful investigation of the impact of write-allocate
policies is also required on other modern processors 
such as AMD- or ARM-based systems.

\Cref{fig:STREAM} shows the bandwidth reported by the STREAM triad  benchmark
on BDW and CLX with (a) and without (b) CoD or SNC enabled. There are three
data sets in each graph: full vectorization with the widest supported
SIMD instruction set and standard stores (noNT), scalar code, and
full vectorization with nontemporal stores (NT).
Note that the scalar and ``noNT'' variants have very similar bandwidth,
which is not surprising since they both cause write-allocate transfers
for an overall code balance of 32\,\BIT. The reported saturated bandwidth of
the ``NT'' variant is a factor of $4/3$ higher because the memory interface
delivers the same bandwidth but the code balance is only 24\,\BIT.\footnote{Note that
  earlier Intel processors like Ivy Bridge and Sandy Bridge could not attain
  the same memory bandwidth with NT stores as without. The difference was small
  enough, however, to still warrant the use of NT stores in performance
  optimization.}

The peculiar shape of the scaling curve with CoD or SNC enabled is a
consequence of the ``compact'' pinning used for the OpenMP threads
(filling the physical cores of the socket from left to right) and the
static scheduling employed by the OpenMP runtime for the STREAM triad
loop. If only part of the second ccNUMA domain is utilized, each of
its active cores will have the same workload as each core on the other
(strongly saturated) domain but more memory bandwidth available. Due
to the implicit barrier at the end of the parallel region, these
``fast'' cores must wait for the slower cores on the other
domain. Hence, each new additional core on the second domain can only
run at the average performance of a core on the first domain, which
leads to the linear scaling. A ``scattered'' pinning strategy would
show only one saturation curve, of course. Note that the 
the available saturated memory bandwidth is independent of the
CoD/SNC setting for the two CPUs under consideration.


\section{Implications for real-world applications}
\label{sec:implications}
In the previous sections we discussed on micro-benchmark
analysis of the two Intel architectures.
In the following we demonstrate how these analysis reflect in
real applications by investigating important kernels such as
DGEMM, sparse matrix-power-vector multiplication and HPCG code.
In order to reflect the practical settings used in HPC runs
we use turbo frequency and switch-off SNC for the following runs.
Measurements which do not follow these settings are clearly mentioned.
Statistical variations are shown whenever the fluctuations are 
bigger than 5\%.

\subsection{DGEMM---Double-precision general matrix-matrix multiplication}
\label{sec:DGEMM}

When implemented correctly, DGEMM is compute-bound on Intel
processors. Each CLX core is capable of executing 32 floating-point operations
(flops) per cycle (8~DP numbers per AVX-512 register, 16\,flops per fused
multiply-add (FMA) instruction, 32\,flops using both AVX-512 FMA units).
Running DGEMM on all twenty cores, the processor specimen from the
testbed managed to sustain a frequency of 2.09\,GHz. The upper limit to
DGEMM performance is thus 1337.6\,GFlop/s.

\begin{figure}[tb]
    \centering
    \hspace*{\fill}
    \includegraphics*[scale=0.70]{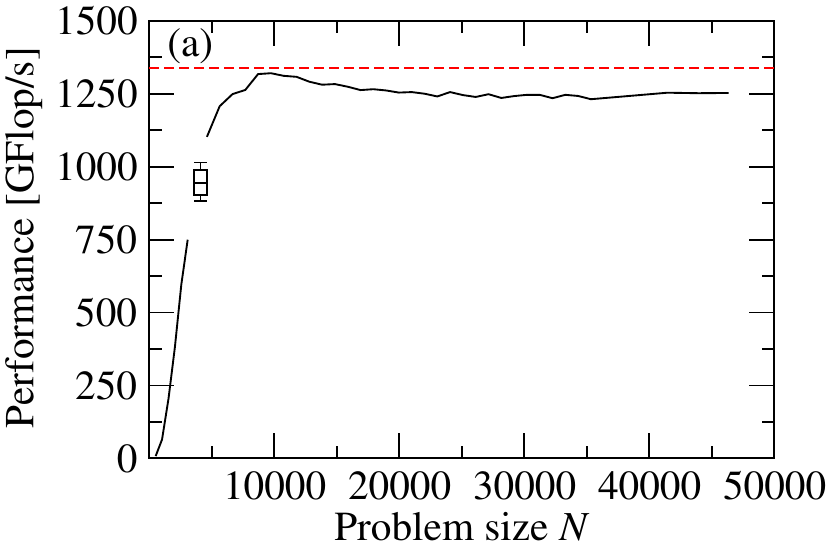}
    \hfill
    \includegraphics*[scale=0.70]{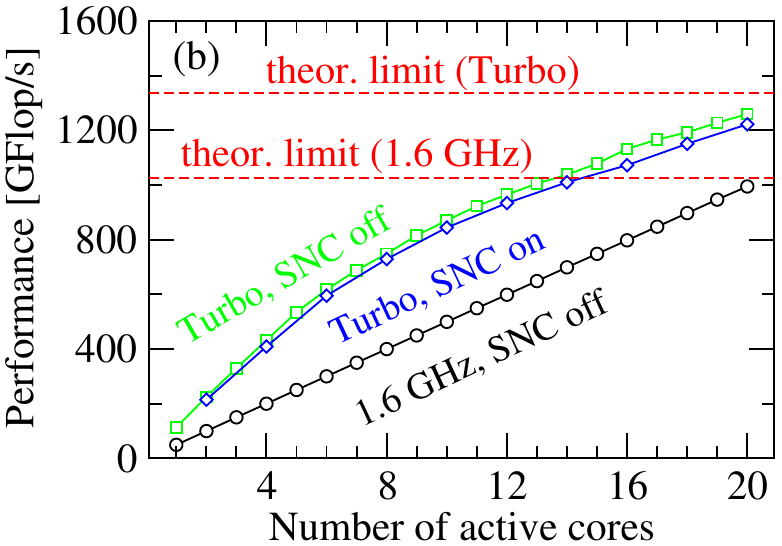}
    \hspace*{\fill}
    \caption{DGEMM performance subject to (a) problem size $N$ and
    (b) number of active cores for $N=40,000$.}
    \label{fig:DGEMM}
\end{figure}

Figure~\ref{fig:DGEMM}a compares measured full-chip DGEMM performance
on CLX in Turbo mode (black line) to theoretical peak performance (dashed red
line). The data indicates that small values of $N$ are not suited to produce
meaningful results. In addition to resulting in sub-optimal performance,
values of $N$ below 10,000 lead to significant variance in measurements, as
demonstrated for $N=4096$ using a box-plot representation (and reproducing the results
from \cite{lmbSANDIA2018}).

Figure~\ref{fig:DGEMM}b shows measured DGEMM performance subject to
the number of active cores.  When the frequency is fixed (in this instance at
1.6\,GHz, which is the frequency the processor guarantees to attain when
running AVX-512 enabled code on all its cores), DGEMM performance
scales practically perfectly with the number of active cores (black line).
Consequently, the change of slope in Turbo mode stems solely from a reduction in
frequency when increasing the number of active cores. Moreover, data shows
that SNC mode is slightly detrimental to performance (blue vs.\ green line).

On Haswell-based processors, a sensitivity of DGEMM performance to
the Uncore frequency could be observed~\cite{Hofmann:2017}: When running cores
in Turbo mode, increasing the Uncore frequency resulted in a decreasing the
share of the processor's TDP available to the cores, which lead them to
decrease their frequency. On CLX this is no longer the case. When running
DGEMM on all cores in Turbo mode results in a clock frequency of
2.09\,GHz---independent of the Uncore frequency. Analysis using hardware
events suggest the Uncore clock is subordinated to the core one: Using the
appropriate MSR (\texttt{0x620}), the Uncore clock can only be increased up to
1.5\,GHz. There are, however, no negative consequences of that limitation.
Traffic analysis in the memory hierarchy indicates that DGEMM is
blocked for the L2 cache, so the Uncore clock (which influences L3 and
memory bandwidth) plays no significant role for DGEMM.


\subsection{SpMPV -- Sparse Matrix-Power-Vector Multiplication}
\label{sec:SpMPV}
The SpMPV benchmark (see \Cref{alg:spmpv}) computes $y=A^px$, where $A$ is a sparse matrix,
as a sequence of sparse matrix-vector products.
The SpMPV kernel is used in a wide range of numerical algorithms like
Chebyshev filter diagonalization for eigenvalue solvers \cite{Chebyshev}, 
stochastic matrix-function estimators used in 
big data applications \cite{stochasticIBM},
and numerical time propagation \cite{timeIntegration}.

The sparse matrix is stored in the compressed row storage (CRS) format
using double precision, and
we choose $p=4$ in our experiments. 
For the basic sparse matrix vector (SpMV) kernel
we use the implementation in Intel MKL 19.0.2.
The benchmark is repeated multiple times to ensure
that it runs for at least one second, so we report the
average performance over many runs. 

For this benchmark we selected  five matrices from the 
publicly available Suite\-Sparse Matrix Collection \cite{UOF}.
The choice of matrices was motivated by some of the 
hardware properties (in particular L3 features) as investigated in previous sections via
microbenchmarks. The details of the chosen matrices are tabulated 
in \Cref{table:SpMPV_mtx}.
The matrices were pre-processed with
reverse Cuthill-McKee (RCM) to attain better data locality;
however, all performance measurements use the pure SpMPV execution
time, ignoring the time taken for reordering. 

\begin{algorithm}[tb]
	\caption{SpMPV algorithm: $y=A^px$}\label{alg:spmpv}
	\begin{algorithmic}[1]
		\State {$double :: A[nnz]$}
		\State {$double :: y[p+1][nrows], x[nrows]$}
		\State {$y[0][*] = x[*]$}
		\For {$i=1:p$}
			\State $y[i][*] = A*y[i-1][*]$
		\EndFor
	\end{algorithmic}
\end{algorithm}

\begin{table}[!tb]
	\caption{\label{table:SpMPV_mtx} Details of the benchmark
	  matrices. $N_{\mathrm{r}}$ is the number of matrix rows,
          $N_{\mathrm{nz}}$ is the number of
	  non-zeros, and $N_{\mathrm{nzr}}=N_\mathrm{nz}/N_\mathrm r$.
          The last column shows the total memory footprint of 
	  the matrix (in CRS storage format) together with the two vectors 
	  required by the benchmark.}
	\centering
	\begin{tabular}{|c|c|S[table-format=7.0, table-space-text-pre=(, table-space-text-post=)]|S[table-format=8.0, table-space-text-pre=(, table-space-text-post=)]|c|c|}
		\toprule
		{Index} & {Matrix name} &  {$N_{\mathrm{r}}$} & {$N_{\mathrm{nz}}$} & {$N_{\mathrm{nzr}}$}  & {size (MB)}  \\
		\midrule
		1 & ct20stif & 52329 & 2698463 & 52 & 32 \\
		2 & boneS01 & 127224 & 6715152 & 53 & 77 \\
		3 & ship\_003 & 121728 & 8086034 & 66 & 93 \\
		4 & pwtk & 217918 & 11634424 & 53 & 134 \\
		5 & dielFilterV3real & 1102824 & 89306020 & 81 & 1024 \\
		\bottomrule
	\end{tabular}
\end{table}

\begin{figure}[!tb]
	\subfloat[L3 scalability]{\label{fig:L3_ct20stif}
		\includegraphics{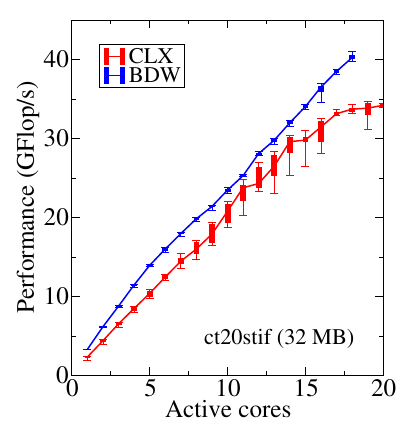}}
	\subfloat[L3 cache replacement]{\label{fig:L3_other_mtx}
		\includegraphics[scale=0.7]{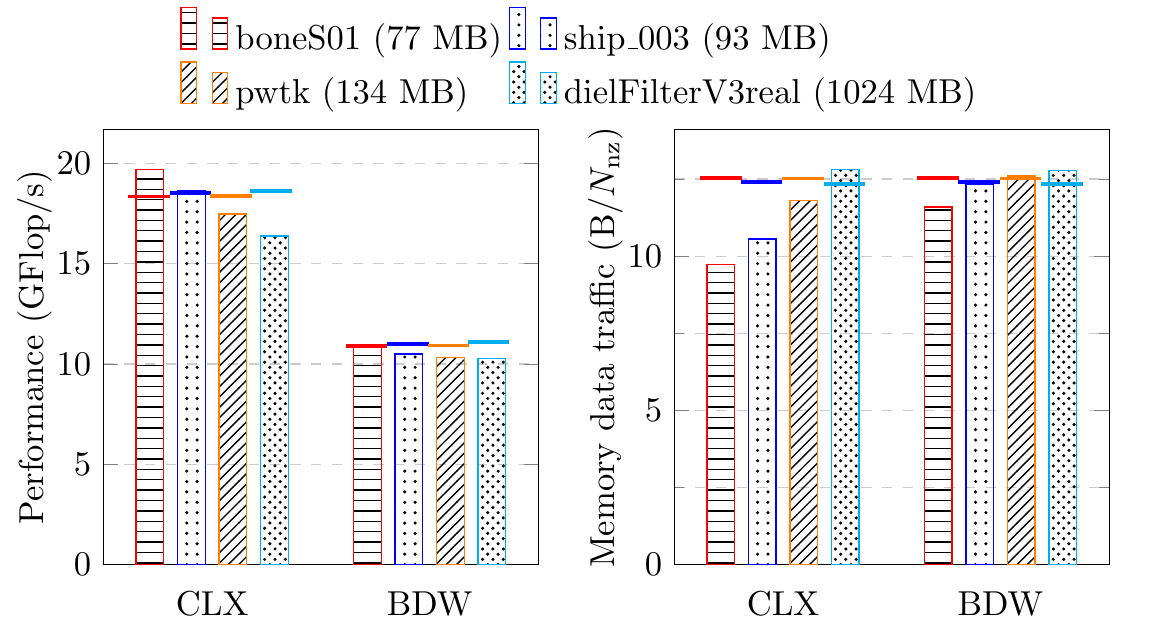}}
	
	\caption{\label{fig:SPMPV}SpMPV benchmark results on
          CLX and BDW. (a)
	  Performance for the \texttt{ct20stif} matrix,
 	  which fits in the L3 cache.
 	  (b) Performance and memory 
 	  data transfer volume for four different 
 	  matrices. Dashed lines mark upper limits  
 	  from a \rlm\ using the saturated load-only
          memory bandwidth.}
\end{figure}


\subsubsection{L3 scalability}
\Cref{fig:L3_ct20stif} shows the scaling performance of the
\texttt{ct20stif} matrix on CLX and BDW. This
matrix is just 30 \MB\ in size and fits easily 
into the caches of both processors.\footnote{Note that
even though the L3 cache of CLX is just 27.5\,\MB\ in size, it is an exclusive victim cache.
The applicable cache size for applications using all cores is
thus 47.5\,\MB, which the aggregate L2/L3 cache size.} 
The L3 bandwidth saturation of CLX as shown
in \Cref{sec:l3_scalability} is reflected by the
performance saturation in the SpMPV benchmark. 
For this matrix, BDW performs better than CLX 
since the sparse matrix kernel is predominantly load bound
and limited by the bandwidth of the load-only micro-benchmark 
(see \Cref{fig:L3_scale_load}). 

Despite its advantage over CLX, the in-cache SpMPV scaling on BDW is 
not linear (parallel efficiency $\varepsilon = 67.5\%$ at all cores),
which stands in contrast with the microbenchmark results in \Cref{fig:L3_scale_load}.
The main reason is that the SpMPV benchmark was run in ``Turbo Mode,''
causing the clock frequency to drop by 25\% when using all the cores
(3.6\,GHZ\ at single core to 2.7\,\GHZ\ at full socket).



\subsubsection{L3 cache replacement policy}
We have seen in \Cref{sec:l3_replacement} that CLX has a more sophisticated
adaptive L3 cache replacement policy, which allows it to extend the 
caching effect for working sets as big as ten times the cache size.
Here we show that SpMPV can profit from this as well.
We choose three matrices that are within five
times the L3 cache size (index 2,3, and 4 in \cref{table:SpMPV_mtx})
 and a moderately large matrix that is 36 times
bigger than the L3 cache (index 5 in \cref{table:SpMPV_mtx}).

\Cref{fig:L3_other_mtx} shows the full-socket performance and memory 
transfer volume for the four matrices. Theoretically, with a 
least-recently used (LRU) policy
the benchmark requires a minimum memory data transfer volume
of \mbox{$12+28/N_{\mathrm{nzr}}$}\,\bytes\
per non-zero entry of the matrix~\cite{RACE}. 
This lower limit is shown in \Cref{fig:L3_other_mtx} 
(right panel) with dashed lines.
We can observe that in some cases the actual memory traffic
is lower than the theoretical minimum, because the L3 cache
can satisfy some of the cacheline requests.
Even though CLX and BDW have almost the same amount of cache,
the effect is more prominent on CLX. On BDW it
is visible only for the \texttt{boneS01} matrix, which is $1.7\times$
bigger than its L3 cache, while on CLX it can be observed
even for larger matrices. 
This is compatible  with the micro-benchmark results in  \Cref{sec:l3_replacement}.
For some matrices the transfer volume is well 
below 12\,\bytes\ per entry, which indicates that not just the vectors
but also some fraction of the matrix stays in cache.

As shown in the left panel of \Cref{fig:L3_other_mtx}, the decrease
in memory traffic directly leads to higher performance. 
For two matrices on CLX the performance is higher than
the maximum predicted by the \Rlm\ 
(dashed line) even when 
using the highest attainable memory bandwidth (load-only).
This is in line with data presented in~\cite{RACE}.


\subsection{HPCG -- High Performance Conjugate Gradient}\label{sec:hpcg}

HPCG\footnote{\url{http://www.hpcg-benchmark.org/}} (High Performance
Conjugate Gradient) is a popular memory-bound proxy application which
mimics the behavior of many realistic \emph{sparse} iterative
algorithms. However, there has been little work to date on analytic
performance modeling of this benchmark.  In this section we analyze
HPCG using the \Rl\ approach.


 
The HPCG benchmark implements a preconditioned conjugate gradient (CG) algorithm with 
a multi-grid (MG) preconditioner. The linear system is derived from a 27-point
stencil discretization, but the corresponding sparse matrix is explicitly stored.
The benchmark uses the two BLAS-1 kernels DOT and WAXPBY
and two kernels (SpMV and MG) involving the sparse matrix.
The chip-level performance of HPCG should thus be governed by the memory bandwidth of
the processor. Since the benchmark prints the \GFS\ performance of all
kernels after a run, this should be straightforward to corroborate.
However, the bandwidths derived from different kernels in HPCG
 vary a lot (see \Cref{table:hpcg_ref}): For the
WAXPBY kernel (\texttt{w[i]=a*x[i]+y[i]}), which has a code balance of 12\,\BF\footnote{The plain WAXPBY kernel
  has a code balance of 16\,\BF\ if a write-allocate transfer must be
  accounted for; however, in HPCG it is called with \texttt{w[]} and \texttt{x[]}
  being the same array, so no write-allocate applies.}, 
the reported performance is  5.14\,\GFS\ on a full socket of BDW. 
On the other hand, for the DOT kernel with a reported code balance of 
8\,\BF\ the benchmark reports a performance of 10.16\,\GFS.
According to the \Rlm\ this translates into memory bandwidths of
61.7\,\GBS\ and 81.3\,\GBS, respectively. The latter value 
is substantially higher than any STREAM value presented for BDW 
in \Cref{fig:STREAM}.
In the following, we use performance analysis and measurements
to explore the cause of this discrepancy, and to check whether the 
HPCG kernel bandwidths are in line with the microbenchmark analysis.

\begin{algorithm}[tb]
	\caption{HPCG \label{alg:HPCG}}
	\begin{algorithmic}[1]
		\While {$k \le iter$ \& $r_{norm}/r_0 > tol$}		
			\State {$z = MG(A,r) $} \Comment{MG sweep}
			\State {$oldrtz = rtz$}
			\State {$rtz = \langle r,z \rangle$} \Comment{DOT}
			\State {$\beta = rtz/oldrtz$}
			\State {$p = \beta*p + z$}	\Comment{WAXPBY}
			\State {$Ap = A*p$}	\Comment{SpMPV}
			\State {$pAp = \langle p,Ap \rangle$} \Comment{DOT}
			\State {$\alpha = rtz/pAp$} 
			\State {$x = x + \alpha*p$} \Comment{WAXPBY}
			\State {$r = r - \alpha*Ap$} \Comment{WAXPBY}
			\State {$r_{norm} = \langle r,r \rangle$}  \Comment{DOT}
			\State {$r_{norm} =sqrt(r_{norm})$}
			\State {$k++$}
		\EndWhile
	\end{algorithmic}
\end{algorithm}

\paragraph{Setup}
For this analysis we use the recent reference variant of HPCG (version 3.1),
which is a straightforward
implementation using hybrid MPI+OpenMP parallelization. 
However, the local symmetric Gauss-Seidel (symGS) smoother used in 
MG has a distance-1 dependency and is not shared-memory 
parallel. The main loop of the benchmark
is shown in \Cref{alg:HPCG}, where $A$ is the sparse matrix 
stored in CRS format.

As the symGS kernel 
consumes more than 80\% of the entire 
runtime, the benchmark is run with pure MPI using one process 
per core. 
The code implements weak scaling across MPI processes;
we choose a local problem size of $160^3$ for a working set of about
1.2\,\GB\ 
per process. The maximum number of CG iteration was set at
25, the highest compiler optimization flag was used 
(see \Cref{tab:testbed}),
and the contiguous storage of sparse matrix data structures was enabled
(\texttt{-DHPCG\_CONTIGUOUS\_ARRAYS}).

\paragraph{Performance analysis of kernels}
We use the \rlm\ to model each of the four kernels separately.
Due to their strongly memory-bound characteristics,
an upper performance limit is given by $P_\mathrm{x}=b_s/C_\mathrm{x}$,
where $b_s$ is the full-socket (saturated) memory bandwidth 
and $C_\mathrm{x}$ is the code balance of the kernel $x$.
 As we have a mixture of BLAS-1 
($N_{\mathrm{r}}$ iterations) and sparse 
($N_{\mathrm{nz}}$ iterations) kernels,
$C_x$ is computed in  terms of \bytes\ required and work done per row of the 
matrix.

The reference implementation contains three DOT kernels (see \Cref{alg:HPCG}).
Two of them require two input vectors to be loaded
(lines 4 and 8 in \Cref{alg:HPCG})
and the other requires just one (norm computation in line 12), 
resulting in a total average code balance
of $C_\mathrm{DOT}=((2\cdot16 + 8)/3)\,\BR=13.33\,\BR$.
All three WAXPBY kernels require one vector to be 
loaded and one vector to be both loaded and stored, resulting in 
$C_\mathrm{WAXPBY}=24\,\BR$. For sparse kernels the total data 
transferred for the inner $N_{\mathrm{nzr}}$ iterations has to be considered.
As shown in \Cref{sec:SpMPV}, the optimal code balance for SpMV
is \mbox{$12+28/N_{\mathrm{nzr}}$} bytes per non-zero matrix entry, 
i.e., $C_\mathrm{SpMV}= (12N_{\mathrm{nzr}}+28)\,\BR$.
For the MG preconditioner we consider only the
finest grid since the coarse grids do not substantially 
contribute to the overall runtime. Therefore the MG consists mainly of one symGS pre-smoothing 
step followed by one SpMV and one symGS post-smoothing step. The symGS
comprises a forward sweep (\verb.0:nrows.) followed by a backward sweep
(\verb.nrows:0.). Both have the same optimal code balance as SpMV,
which means that the entire MG operation has a code balance of five times that
of SpMV: $C_\mathrm{MG} = 5C_\mathrm{SpMV}$.

The correctness of the predicted code balance can be
verified using performance counters. 
We use the \likwidperfctr\ tool to count the number of 
main memory data transfers for each of the kernels.\footnote{See \url{https://github.com/RRZE-HPC/likwid/wiki/TestAccuracy} for validation
of the data groups.} 
\Cref{table:hpcg_ref} summarizes the predicted and measured 
code balance values for full-socket execution along with the reported performance
and number of flops per row for the four kernels in HPCG. Except for DDOT,
the deviation between predicted and measured code balance is less than 10\%. 
 
\paragraph{MPI desynchronization}
Surprisingly, DDOT has a measured code balance that is lower than the
model, pointing towards caching effects.
However, a single input vector for DDOT has a size of 560\,\MB, which
is more than ten times the available cache size. As shown in
\Cref{sec:l3_replacement}, even CLX 
is not able to show any significant caching effect
with such working sets.
Closer investigation revealed \emph{desynchronization} of MPI processes
to be the reason for the low code balance:
In \Cref{alg:HPCG} we can see that some of the DOT kernel can
reuse  data from previous kernels. For example, the last DOT
(line 12) reuses the $r$ vector form the preceding WAXPBY. 
Therefore, if MPI processes desynchronize such that only some
of them are already in DOT while the others are in different kernels	
(like WAXPBY), then the processes in DOT can reuse the data, while
the others just need to stream data as there is no reuse. This
is possible since there are no explicit synchronization points
between previous kernels and DOT.

\begin{figure}[!tb]
	\subfloat[BDW]{\label{fig:hpcg_scaling_bdw}\includegraphics[scale=1]{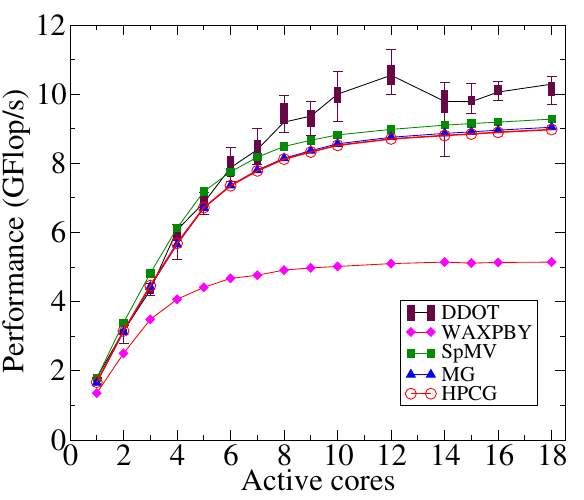}}
	\subfloat[CLX]{\label{fig:hpcg_scaling_clx}\includegraphics[scale=1]{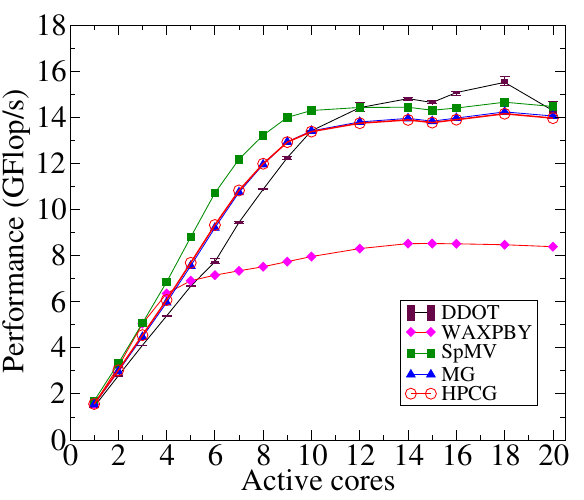}}
	\caption{\label{fig:hpcg_scaling} Performance of different
	  kernels in the HPCG benchmark (reference implementation)
     as a function of active cores.}
\end{figure}

This desynchronization effect is not predictable and will vary between runs
and machine as can be observed by the significant performance fluctuation 
of DOT in \Cref{fig:hpcg_scaling}. 
To verify our assumption we added barriers 
before the DOT kernels, which caused the measured 
$C_\mathrm{DOT}$ to go up to 13.3\,\BR, matching the expected value.
The desynchronization effect clearly shows  the
importance of analyzing statistical fluctuations and deviations
from performance models. Ignoring them can easily lead to false conclusions
about hardware characteristics and code behavior.

\begin{table}[!tb]
	\centering
	\caption{Summary of the \rl\ performance model parameters 
		and measurements for HPCG kernels.
		Predicted and measured values for code balance and performance
		are shown in columns three to six. 
		The last two columns compare the 
		predicted and measured performance of the entire solver.
	}
	\label{table:hpcg_ref}
	\begin{tabular}{|l|l|c|c|c|c|c|c|c|c|}
		\toprule
		\multirow{3}{*}{Arch} & \multirow{3}{*}{Kernels} &  \multicolumn{2}{c|}{Code-balance ($C_\mathrm{x}$)} & \multicolumn{2}{c|}{Performance ($P_\mathrm{x}$)} &  \multirow{1}{*}{Flops} & \multirow{1}{*}{Calls} & \multicolumn{2}{c|}{HPCG perf.}\\
		\cline{3-6}
		\cline{9-10}
		& & {Predicted} & {Measured} &  {Predicted} & {Measured} & ($F_\mathrm{x}$) & ($I_\mathrm{x}$) & {Predicted} & {Measured}\\
		\cline{3-6}
		\cline{9-10}
		\cline{7-7}
		& & {\BR} & {\BR} & {\GFS} & {\GFS} & {\FR} & & {\GFS} & {\GFS}\\
		\midrule
		\multirow{4}{*}{BDW}
		& DDOT & 13.30 & 11.13 & 10.23 & 10.16 & 2 & 3  & \multirow{4}{*}{10.27} & \multirow{4}{*}{8.98}\\
		& WAXPBY & 24.00 & 24.11 & 5.67 & 5.14 & 2 & 3  &  & \\
		& SpMV & 352.00 & 385.61 & 10.43 & 9.28 & 54 & 1  &  & \\
		& MG & 1760.00 & 1952.09 & 10.43 & 9.04 & 270 & 1  &  & \\
		\midrule
		\multirow{4}{*}{CLX}
		& DDOT & 13.30 & 12.68 & 17.29 & 14.34 & 2 & 3  & \multirow{4}{*}{17.37} & \multirow{4}{*}{13.95}\\
		& WAXPBY & 24.00 & 24.02 & 9.58 & 8.39 & 2 & 3  &  & \\
		& SpMV & 352.00 & 382.68 & 17.64 & 14.46 & 54 & 1  &  & \\
		& MG & 1760.00 & 1944.31 & 17.64 & 14.05 & 270 & 1  &  & \\
		\bottomrule
	\end{tabular}
\end{table}

\paragraph{Combining kernel predictions}
Once the performance predictions for individual kernels are in place, we can 
combine them to get a prediction of the entire HPCG.
This is done by using a time-based formulation of the \rlm\ and linearly
combining the predicted kernel runtimes based on their call counts.
If $F_\mathrm{x}$ is the number of flops per row and $I_\mathrm{x}$ the 
number of times the kernel $x$ is invoked,
the final prediction is
\begin{align}
	T_\mathrm{HPCG} &= \sum_{x}^{}I_\mathrm{x}T_\mathrm{x} \quad \forall
	x \in \{\mathrm{DOT, WAXPBY, SpMV, MG}\}~,\\
	\text{where~~} T_\mathrm{x} &=  F_\mathrm{x}N_{\mathrm{r}}/P_\mathrm{x}~. 
\end{align}
\Cref{table:hpcg_ref} gives an overview of $F_\mathrm{x}$, $I_\mathrm{x}$, 
and $C_\mathrm{x}$ for different kernels and compares the predicted and 
measured performance on a full socket.
The prediction is consistently higher than the model because we used
the highest attainable bandwidth for the \rlm\ prediction.
For Intel processors this is the load-only bandwidth $b_\mathrm S=115\,\GBS$
(68\,\GBS) for CLX (BDW), which is
approximately 10\% higher than the STREAM values (cf. \Cref{sec:multicore}).
\Cref{fig:hpcg_scaling} shows the scaling
performance of the different kernels in HPCG. The typical saturation
pattern of memory-bound code can be observed on both
architectures.
\section{Conclusions and Outlook}
\label{sec:summary}
Two recent, state of the art generations
of Intel architectures have been analyzed: Broadwell EP and Cascade Lake SP.
We started with a basic microarchitectural
study concentrating on data access.
The analysis showed that our benchmarks were able to obtain 85\% of the 
theoretical limits and for the first time demonstrated 
the performance effect of the Intel's newly designed shared L3 victim cache.
 During the process of microbenchmarking
 we also identified the importance of selecting proper benchmark tools
 and the impact of various hardware, software, and OS settings, thereby
 demonstrating the necessity for detailed documentation. 
We further demonstrated that the  observations made in microbenchmark analysis 
are well reflected in real-world application scenarios. To this extent we
investigated the performance characteristics of DGEMM, sparse matrix-vector multiplication,
and HPCG.
For the first time a \rl\ model of HPCG and its components
was established and successfully validated for both architectures.
Performance modeling was used as a guiding tool
throughout this work to get deeper insight and explain anomalies.

Future work will include investigation of benchmarks for
random and latency-bound codes along with the development of
 suitable performance models. 
The existing and further upcoming wide range of architectures will bring 
more parameters and benchmarking challenges, which will be 
very interesting and worthwhile to investigate.




\ifblind
\else
\subsection*{Acknowledgments}
We are indebted to Thomas Zeiser and Michael Meier (RRZE) for
providing a reliable benchmarking environment.
This work was partially funded via the ESSEX project in the DFG
priority programme 1648 (SPPEXA) and by the German Ministry of Science and
Education (BMBF) under project number 01IH16012C (SeASiTe).
\fi

\bibliographystyle{splncs03}
\bibliography{pub} 


\end{document}